\documentclass[prl,twocolumn,twoside,superscriptaddress]{revtex4}

\usepackage{amsmath}
\usepackage{graphicx}

\begin{document}




\title{Dynamics of channel incision in a granular bed driven by
  subsurface water flow}

\author{Alexander E.~Lobkovsky}
\affiliation{Department of Earth, Atmospheric and Planetary Sciences,
  MIT, 77 Massachusetts Avenue, Cambridge, MA 02139}

\author{Braunen Smith}
\author{Arshad Kudrolli}
\affiliation{Department of Physics, Clark University, Worcester, MA 01610}

\author{Daniel H.~Rothman}
\affiliation{Department of Earth, Atmospheric and Planetary Sciences,
  MIT, 77 Massachusetts Avenue, Cambridge, MA 02139}



\begin{abstract}
  We propose a dynamical model for the erosive growth of a channel in
  a granular medium driven by subsurface water flow.  The model is
  inferred from experimental data acquired with a laser-aided imaging
  technique.  The evolution equation for transverse sections of a
  channel has the form of a non-locally driven Burgers equation.  With
  fixed coefficients this equation admits an asymptotic similarity
  solution.  Ratios of the granular transport coefficients can
  therefore be extracted from the shape of channels that have evolved
  in steady driving conditions.
\end{abstract}

\maketitle

One of the salient features of fluvial erosion is the formation and
growth of channels~\cite{dietrich93:_channel}.  The channels act to
focus the flow of water, either over land~\cite{horton45}, or,
ubiquitously but less commonly appreciated, beneath the
surface~\cite{dunne90}.  Here we address the dynamics of erosion
driven by subsurface flow.  Whereas flow through the subsurface is
well characterized by Darcy's
law~\cite{scheidegger60:_book,bear72:_book}, channel formation and
growth resulting from seepage flows out of the surface is relatively
poorly understood~\cite{schorghofer04:_spont}.  This erosive process
is important not only because of its relevance to enigmatic features
of natural
landscapes~\cite{howard88:_seepage,laity85:_sapping,schumm95:_florida,orange94:_spacing,higgins82:_sapping,schorghofer04:_spont,wentworth28:_hawaii,kochel86:_hawaii,malin00:_seepage_mars},
but also because it raises fundamental questions concerning the
continuum mechanics of wet sand
\cite{daerr:065201,huang05:_flow_wet,fournier05:_mech_wet,tegzes03:_wet_avalanches,schulz03:_shear_wet,jain04:_fluid_layer,jain01:_self,zhou05:_shocks}.



In this Letter we construct a theoretical model of the erosive
dynamics of an isolated channel.  The model is inferred from data
acquired with a laser-aided imaging technique.  In our experiment we
grow a single channel under controlled driving conditions and acquire
maps of the evolving granular surface that are highly resolved in
space and time.  We find that transverse sections of a channel evolve
according to a non-locally driven Burgers equation.  Under steady
driving conditions the evolution equation admits a similarity
solution, thereby relating erosive transport coefficients to channel
geometry.


\begin{figure}[htbp]
  \centering
  \includegraphics[width=8cm]{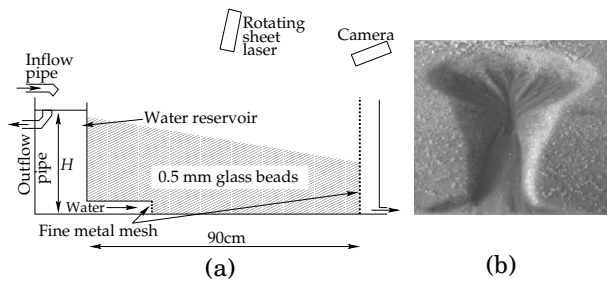}
  \caption{(a) Schematic of our experimental setup.  (b) Example of an
    isolated channel observed after 90 minutes of subsurface water
    flow driven growth.  A small initial channel is incised on the
    surface to fix the location and number of channels.  Similar
    shapes are observed if channels form spontaneously.
    \label{fig:expsetup}}
\end{figure}

Figure~\ref{fig:expsetup} shows a schematic of the experiment, aspects
of which are described in more detail elsewhere
\cite{schorghofer04:_spont,lobkovsky04:_thresh}.  Water enters beneath
a pile of identical glass beads through a fine mesh and exits at the
foot of the pile through the same kind of mesh.  The water flux is
controlled by the height $H$ of the water column in a reservoir behind
the pile.  The slope of the initial sandpile as well as the water
column height are the control variables of the experiment.  Since the
water table is convex upward in this
geometry~\cite{lobkovsky04:_thresh}, we are able to finely control the
amount of water on the surface of the granular pile.
Refs.~\cite{schorghofer04:_spont} and \cite{lobkovsky04:_thresh}
describe the phenomenology of the pattern formation in this setup and
quantify the transitions between various modes of granular flow:
surface flow (responsible for the formation of the channel network),
slumping (bulk frictional instability) and fluidization.

Here we focus on the late-stage evolution of an isolated channel that
grows from a short initial channel of triangular crossection.  A
scanning laser imaging technique allows us to measure the evolving
height of the sandpile with sub-grain resolution in space and
one-minute resolution in time.  A laser sheet scans the surface while
a digital camera acquires images from an oblique angle.  The height of
the surface is then extracted from an image of the intersection of the
laser sheet with the granular surface.

The water level $H$ is set below the threshold for erosion outside the
channel but above the threshold for erosion in the channel.  An
example of a channel grown from an initial channel with a triangular
crossection is shown in Fig.~\ref{fig:expsetup}(b).  We find that the
late stage morphology of the channel is insensitive to the exact
initial condition as long as the initial incision is sufficiently
deep.

\begin{figure*}[tbp]
  \centering
  \includegraphics[width=14cm]{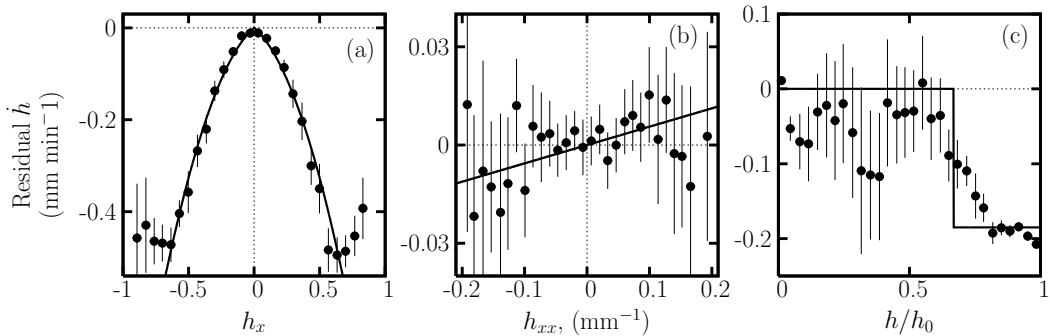}
  \caption{Residual erosion rate at $t = 39$ min plotted against (a)
    the slope, (b) the curvature, and (c) the fractional channel
    depth.  Each symbol represents an average over several hundred
    data points.  To obtain the residual erosion rate we subtract from
    the measured $\dot h$ the terms on the right hand side of
    Eq.~\eqref{eq:erosion_rate} which depend on the variables other
    than the one against which the erosion rate is being plotted.
    Lines are drawn using the parameters extracted by a least squares
    fit of the unbinned data to equation \eqref{eq:erosion_rate}.
    Error bars represent two standard deviations of the mean.
    \label{fig:dyn_fit}}
\end{figure*}

We seek an effective description of the channel's evolution in terms
of the surface height $h(x,y,t)$ measured from the uneroded surface.
In other words, we seek an expression for the erosion rate $\dot h
\equiv \partial h/\partial t$ in terms of $h$, its spatial
derivatives, and possibly spatial integrals.  (Here $y$ is the
downslope axis and $x$ is the axis transverse to the channel).  This
approach is reasonable because the relaxation of the water flux toward
steady state is fast compared to the erosion rate.  Thus the water
flux is a functional of the slowly changing shape of the channel.

The erosion rate is a function of the local water and granular fluxes.
In steady state these fluxes are functionals of the global shape of
the granular pile.  We seek to reduce the global dependence to a
single scalar by arguing that, because the channel evolves in a
roughly self-similar manner, the water fluxes (bulk as well as
surficial) are functions of the local topography and a time-varying
scale factor.  We choose the channel depth $h_0(y)$ as this scale
factor ($h_0$ corresponds to height of the deepest point of the
crossection). We have thus reduced the problem to finding the
dependence of the erosion rate on the local topography and the overall
global scale factor.

We make one more simplification: we consider solely the evolution of
transverse sections of the channel, and therefore express the erosion
rate $\dot h \equiv \partial h/\partial t$ through $h$ and its
derivatives with respect to the transverse coordinate $x$ only.
Variation of the water flow in the downslope $y$-direction is
accounted for via $y$-dependent coefficients.  This approximation is
reasonable everywhere except the top portion of the channel's head
where the downslope gradient varies rapidly.  Grains and microscopic
avalanches enter and leave a given channel transect.  Projected onto
this transect, the transport of height $h$ is no longer volume
conserving.

Thus, the erosion rate $\dot h$ is assumed to be a function of the
fractional depth $h/h_0$, the transverse slope $h_x = \partial
h/\partial x$, and the curvature $h_{xx} = \partial^2 h/\partial x^2$.
Our approach is to measure these quantities for data in a window of
duration $\Delta t$ in time and length $\Delta y$ in the downslope
coordinate and to fit the resulting data cloud via a least squares
method to the form
\begin{equation}
  \label{eq:erosion_rate}
  \dot h = \nu h_{xx} - \delta |h_x| - \lambda h_x^2 -
  \mu \, \Theta(h/h_0 - f),
\end{equation}
where $\Theta(x)$ is the Heaviside step function.  The empirical
constants $\mu$, $\nu$, $\lambda$, $\delta$, and $f$ are functions of
time $t$ and downslope coordinate $y$.  They encode the microscopic
properties of the grain dynamics as well as the strength of the
driving water flow.  The diffusion constant $\nu$ reflects the rate of
smoothing of local perturbations.  The last term on the right hand
side represents driving due to the seeping water.  We assume that
driving is constant below a fraction $f$ of the total channel depth
and null above this fractional depth.  The second and third terms on
the right hand side of \eqref{eq:erosion_rate} are advective.  We
hypothesize that perturbations are advected only up the slope, similar
to Ref.~\cite{boutreux98:_surf_flows1}.  Therefore, the non-analytic
term $\delta |h_x|$ corresponds to advection of perturbations with
velocity $\delta$ independent of slope, whereas $\lambda h_x^2$
corresponds to advection with velocity $\lambda h_x$, which grows
linearly with slope.

Figure \ref{fig:dyn_fit} illustrates the fit of the data to the model.
As shown in Fig.~\ref{fig:dyn_fit}(a), the slope dependent terms make
the largest contribution into the erosion rate.  Data points with
slopes smaller than 0.7, comprising over 95\% of all data, have been
used in the fit.  A discrepancy between the model and the data occurs
for larger slopes where the erosion rate saturates.  A measurable
contribution to the erosion rate due to diffusion is shown in
Fig.~\ref{fig:dyn_fit}(b).  The extracted positive granular diffusivity
$\nu$ is statistically significant.  The driving, depth-dependent
term, shown in Fig.~\ref{fig:dyn_fit}(c), is approximated by a step
function although the data show a more gradual transition.  We expect
the width of the transition region to remain constant as the channel
grows.  Thus the step function approximation should improve with time.

\begin{figure}
  \centering
  \includegraphics[width=5cm]{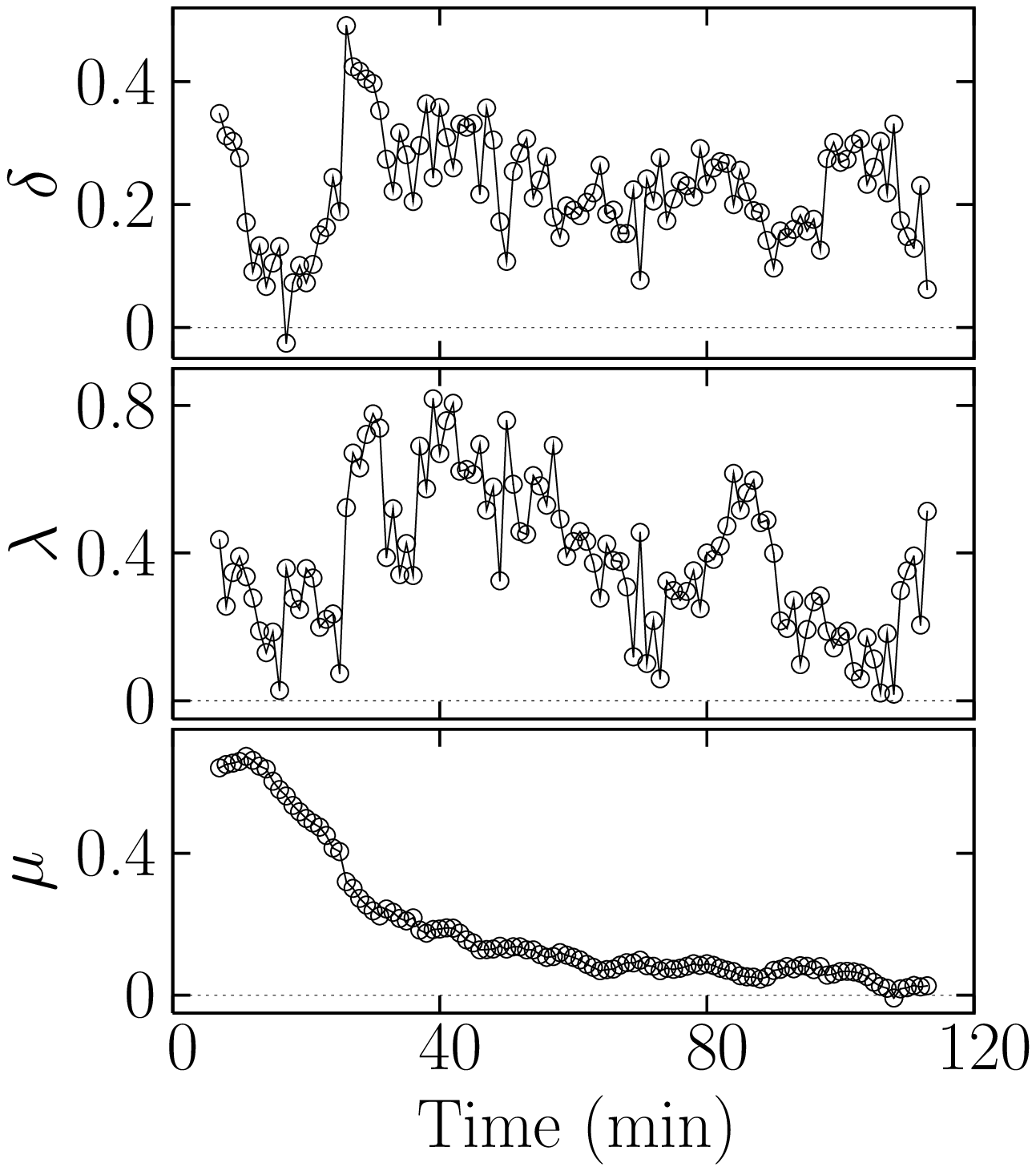}
  \caption{Example of the evolution of the coefficients $\mu$,
    $\lambda$ and $\delta$ (measured in mm/min) extracted using
    Eq.~\eqref{eq:erosion_rate} from the data using taken near a fixed
    transect illustrated in Fig.~\ref{fig:comparison}.  The head of
    the channel arrives to this transect approximately one minute
    after the start of the experimental run.  \label{fig:parameters}}
\end{figure}

We extract the time-dependent coefficients in
Eq.~\eqref{eq:erosion_rate} near a transect fixed in the lab frame.
Fig.~\ref{fig:parameters} shows the extracted parameters for a
transect initially above the pre-dug channel.  As the channel head
migrates past this transect, the driving $\mu$ peaks and declines.
The values of advection speeds $\delta$ and $\lambda$ also decrease
since they are related to the driving.  The diffusion coefficient $\nu
= 0.039 \pm 0.033$ mm$^2$/min and the fractional driving $f = 0.52 \pm
0.14$ fluctuate around their respective means which are roughly
time-independent.

\begin{figure}
  \centering
  \begin{tabular}[t]{cc}
    \includegraphics[height=3.5cm]{contours} &
    \includegraphics[height=3.5cm]{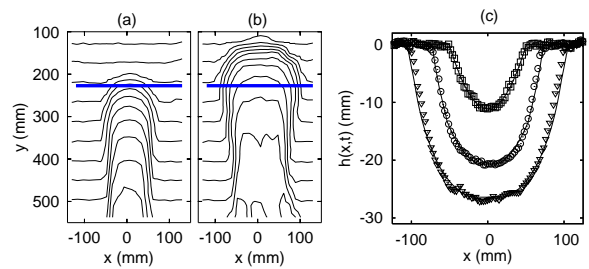}
  \end{tabular}
  \caption{Contour plots of the laser height data at (a) $t = 15$ min
    and (b) 115 min after the start of the water flow.  The thick line
    shows the location of the modeled transect.  The contour interval
    is 5 mm.  (c) Same transect at $t = $ 15, 45, and 115 min compared
    to the prediction of Equation~\eqref{eq:erosion_rate}.  The
    smoothed shape of the transect at $t = 15$ min was used as the
    initial condition.  Actual time dependent parameters presented in
    Fig.~\ref{fig:parameters} were used.
    \label{fig:comparison}}
\end{figure}

In Fig.~\ref{fig:comparison}(a) and (b) we indicate in a contour plot
of the channel the location of the transect for which the parameters
in Fig.~\ref{fig:parameters} are computed.
Fig.~\ref{fig:comparison}(c) compares the measured shapes of this
transect with the shapes evolved via Eq.~\eqref{eq:erosion_rate} with
the time-dependent coefficients.  Agreement indicates that the right
hand side of Eq.~\eqref{eq:erosion_rate} captures the essential
features of the erosion rate in the channel geometry.  The rate of
advance of the channel sidewalls and the receding rate of channel
bottom are determined by different combinations of the parameters.
The match of the evolved shapes to the data shows that both rates are
in accord with the experiment.

In our experiment the coefficients, intimately related to the geometry
of the water flow, change markedly during the evolution of the channel
as illustrated in Fig.~\ref{fig:parameters}.  It is conceivable,
however, that in other geometries the transport coefficients are
approximately constant over a long time.  It is therefore appropriate
to examine the long-time behavior of Eq.~\eqref{eq:erosion_rate} with
constant coefficients.  The rest of the paper is devoted to the
asymptotic calculation.

The main result is that the non-local driving term in
Eq.~\eqref{eq:erosion_rate} allows for an asymptotically self-similar
growing channel shape.  To see that let us define a scale factor
$\ell(t)$ and the shape $Y(\xi,t)$, which is a function of the scaled
coordinate $\xi = x/\ell(t)$ via a transformation $h(x,t) = \ell(t) \,
Y(\xi, t)$.  Further progress results from a formal expansion
\begin{subequations}
  \label{eq:expansion}
  \begin{eqnarray}
    \label{eq:y_expand}
    Y(\xi,t) &=& Y_0(\xi) + \frac{1}{t} \, Y_1(\xi) + \ldots, \\
    \label{eq:ell_expand}
    \ell(t) &=& t + \ell_1 \ln t + \ldots.
  \end{eqnarray}
\end{subequations}
The zeroth order shape $Y_0(\xi)$ is independent of time.  Thus, if
the expansion \eqref{eq:expansion} converges, the shape converges
(albeit slowly as $1/t$) to a similarity solution $Y_0$.  A
substitution of the expansion \eqref{eq:expansion} into
\eqref{eq:erosion_rate} yields, at the lowest order,
\begin{equation}
  \label{eq:Y_0_eq}
    Y_0 - \xi Y_0' = -\beta Y_0' - \alpha (Y_0')^2 -
    \Theta\left(\frac{Y_0}{Y_0(0)} - f\right),
\end{equation}
where primes denote differentiation with respect to $\xi$.  We scaled
lengths by $a = \nu/\mu$ and time by $\tau = \nu/\mu^2$ and defined
$\beta = \delta/\mu$ and $\alpha = \lambda/\mu$.  Note that the
diffusive term does not enter at the lowest order.  The physical
reason for this is that, as we show below, diffusion is important on a
fixed length scale.  As the channel grows larger than this diffusive
length, the $h_{xx}$ term in \eqref{eq:erosion_rate} becomes
negligible compared to the advection and the driving terms.
 
The solution to \eqref{eq:Y_0_eq} must be symmetric with respect to
$\xi \rightarrow -\xi$ and smooth everywhere except at the fractional
driving point $\xi_0$ where $Y_0(\xi_0) = f Y_0(0)$.  It turns out
that there exists a one-parameter family of similarity solutions which
satisfy these criteria.  These solutions are constructed as follows.
In the driven region any piece of the parabola
\begin{equation}
  \label{eq:parabola}
  Y_0(\xi) = \frac{1}{4\alpha}\left(\xi^2 - 2\beta\xi + \beta^2 -
    4\alpha\right),
\end{equation}
and a tangent line to this parabola at a point $\xi_1$,
\begin{equation}
  \label{eq:tangent}
  Y_0(\xi) = \frac{1}{4\alpha}\left(2\xi (\xi_1 - \beta) + \beta^2 -
    \xi_1^2 - 4\alpha\right), 
\end{equation}
are solutions to \eqref{eq:Y_0_eq}.  In the undriven region, the
trivial solution $h = 0$ as well as tangents to the parabola
\begin{equation}
  \label{eq:parabola-undriven}
  Y_0(\xi) = \frac{1}{4\alpha}\left(\xi^2 - 2\beta\xi\right)
\end{equation}
are solutions.  A smooth solution symmetric around $\xi=0$ is therefore
constructed from four pieces:
\begin{equation}
  \label{eq:zeroth_solution}
  Y_0 =
  \begin{cases}
    -1, & 0 < \xi < \xi_b, \\
    -1 + \frac{(\xi - \xi_b)^2}{4\alpha}, & \xi_b
    < \xi < \xi_1, \\
    -1 + \frac{(\xi_1 - \xi_b)^2}{4\alpha} + \frac{(\xi_1 - \xi_b)(\xi
      - \xi_1)}{2\alpha}, & \xi_1 < \xi  < \xi_0\\
    -f + s_0(\xi - \xi_0), & \xi_0 < \xi < \xi_e.
  \end{cases}
\end{equation}
The first piece is the flat bottom of the channel tangent to the
parabola \eqref{eq:parabola} at its apex $\xi_b = \beta$, $Y_0 = -1$.
The second piece is part of the parabola itself.  The third piece is
another tangent to parabola \eqref{eq:parabola} at a point $\xi_1$
such that $\beta \leq \xi_1 \leq \beta + \sqrt{4\alpha(1 - f)}$.  The
fourth piece is the tangent of slope $s_0$ to the undriven parabola
\eqref{eq:parabola-undriven}.  Note that $Y_0$ is not smooth at
$\xi_0$ (driving) and $\xi_e$ (channel edge).

It turns out that the diffusive term, though not present at zeroth
order, acts to select a unique member from the one-parameter family of
similarity solutions.  The selection mechanism, verified by numerical
methods, is unclear to us at this time.  The selected similarity
solution is one in which the second tangent to the parabola
\eqref{eq:parabola} is missing, i.e., $\xi_1$ assumes its upper limit.
The slope of the sidewall of the channel in the undriven region is
then simply
\begin{equation}
  \label{eq:simple_flank-slope}
  s_0 = \frac{1}{\sqrt{\alpha}}(1 + \sqrt{1 - f}).
\end{equation}
The expressions for $\xi_0$ and $\xi_e$ are simple as well:
\begin{equation}
  \label{eq:xi_0_and_xi_e}
  \xi_0 = \beta + \sqrt{4\alpha(1 - f)}, \quad \xi_e = \xi_0 + f/s_0
\end{equation}

We remark that the asymptotic shape consisting of a flat bottom,
curved parabolic flanks in the driven region followed by straight
sidewalls can be used successfully to fit channel crossections which
evolve in non-steady conditions.  However, the ratios of parameters
extracted by such an asymptotic fit can deviate greatly from the
parameters extracted by the fit of the equation
\eqref{eq:erosion_rate} to the data cloud, because the shape of the
channel at the time of the fit retains a memory of its prior dynamical
state.

Besides selecting the unique self-similar shape, the diffusive term
also acts to smooth slope discontinuities at $\xi_0$ and $\xi_e$.  An
exact expression for this smoothing can be obtained at the channel's
edge $\xi_e$.  In the vicinity of this point, the slope $s = h_x$ is a
hyperbolic tangent
\begin{equation}
  \label{eq:smooth_corner}
  s(x,t) = \frac{s_0}{2}\left(1 - \tanh{\frac{s_0 \alpha}{2}(x -
      v t)}\right),
\end{equation}
moving with velocity $v = s_0 \alpha + \beta$ and smooth on a scale
\begin{equation}
  \label{eq:R}
  R = \frac{2\nu}{\sqrt{\lambda\mu}(1 + \sqrt{1 - f})}.
\end{equation}
As we mentioned above, diffusion acts on a fixed scale $R$.  The
smoothing of the kink at $\xi = \xi_0$ is probably related to the
selection of the unique similarity solution.

Guided by an experiment, we have developed a phenomenological model of
granular dynamics in a transect of a channel eroded by subsurface
fluid flow.  Precise, time resolved measurements of the height of the
eroding sandpile allow us to test the validity of the model.  The
erosion rate is a function of the local topography and an overall
scale factor.  We expect this approximation to be good as long as the
channel's shape remains roughly self-similar.  There are five
parameters in the model which in our experiment vary with time.  These
parameters encode the geometry of the water flow as well as the
features of the microscopic granular flow.  In a different water flow
geometry these parameters can remain roughly constant.  In this case
the height evolution equation \eqref{eq:erosion_rate} admits a
self-similar solution and the three dimensionless parameters $\alpha$,
$\beta$ and $f$ dictate the self-similar shape.  Conversely, if a
shape is known to have resulted from evolution with roughly constant
parameters, then $\alpha$, $\beta$ and $f$ can be extracted from the
static shape.  Given an independent measure of the erosion rate $\mu$
at the bottom of the channel, as well as the measurement of the
smoothing length scale $R$ at the channel's edge, all five parameters
can be recovered.  Such a procedure should be useful for the analysis
of natural channels formed by subsurface fluid flow.

This work is supported by DOE grants DE--FG02--02ER15367 at Clark
University and DE--FG02--99ER15004 at MIT.

\bibliography{/home/leapfrog/Manuscripts/all}
\end{document}